\documentclass[10pt,twocolumn,aps,showpacs,superscriptaddress]{revtex4-1}
\newcommand{\be}{\begin{equation}}
\newcommand{\ee}{\end{equation}}
\newcommand{\br}{\begin{eqnarray}}
\newcommand{\er}{\end{eqnarray}}
\usepackage{times}
\usepackage{color}
\usepackage{graphicx}
\usepackage{bm}
\usepackage{amsmath}
\usepackage{amsfonts}
\usepackage{amsthm}
\usepackage{amssymb}
\usepackage{mathrsfs}
\begin{document}
\title{Minkowski structure for purity and entanglement of Gaussian bipartite states}
\author{Marcos C. de Oliveira}
\email{marcos@ifi.unicamp.br}
\affiliation{Instituto de F\'{\i}sica Gleb Wataghin, Universidade Estadual de Campinas, 13083-859, Campinas, S\~{a}o Paulo,
Brazil}
\author{Fernando Nicacio}\affiliation{Instituto de F\'{\i}sica Gleb Wataghin, Universidade Estadual de Campinas, 13083-859, Campinas, S\~{a}o Paulo,
Brazil}
\author{Salomon S. Mizrahi}
\affiliation{Departamento de F\'\i sica, CCET, Universidade Federal de S\~ao Carlos,
 Via Washington Luiz Km 235, S\~ao Carlos, 13565-905, SP, Brazil.}
\date{\today}
%
\begin{abstract}
The relation between the symplectic and  Lorentz groups is explored to 
investigate entanglement features in a two-mode bipartite Gaussian state.
We verify that the correlation matrix of arbitrary Gaussian states can be associated 
to a hyperbolic space with a Minkowski metric, which is divided in two regions - \emph{separablelike}, 
and \emph{entangledlike}, in equivalence to timelike and spacelike in special relativity. 
This correspondence naturally allows the definition of two insightful 
invariant squared distances measures - 
one related to the purity and another related to amount of entanglement. 
The second distance allows us to define a measure for entanglement in terms of the invariant interval between the given state and its closest separable state, given in a natural manner without the requirement of a minimization procedure.\end{abstract}
%
\maketitle
%
{\textit{Introduction.}}
The symplectic group is isomorphic to the structure of the 
Lorentz and de Sitter groups, as was firstly pointed out
by Dirac himself in his famous 3 + 2 de Sitter group article \cite{Dirac}. 
 In fact, all Gaussian light field states embody the symplectic structure \cite{kim}, 
as has been explored in the implementation of several features such as quadrature squeezing and quantum entanglement.
A particularly important separability criterion, based on the symplectic structure of Gaussian 
states (GS), 
was given by Simon \cite{simon}, as an extension for continuous variables of the Peres-Horodecki positivity under partial transposition (PPT) criterion \cite{peres,horodecki}. 
It is remarkable that positive maps
can actually be associated to a hyperbolic geometry displaying formal similarity with the spacetime manifold of special relativity.
This connection was reported earlier \cite{hss1,hss2} for two-qubit systems 
where the concept of hyperbolic squared distance was introduced as a measure of entanglement, within a compact support in contrast with the space-time manyfold.
The relation of the invariants of the Lorentz group, namely space-time
squared intervals, with transformations and entanglement properties of GS 
seems to us quite advantageous to be seen from a geometric perspective.
 
In this paper we give a geometrical picture of the separability bound for two-mode bipartite
GS in terms of a hyperbolic geometry having a Minkowski metric, and explore the formal 
similarities between purity and entanglement properties with some familiar concepts in theory of 
relativity. The advantage of such an approach is made clear for the 
definition of distances related to entanglement and purity measures in terms 
of invariant intervals, which do not rely 
on some optimization procedure as usual \cite{paulina2,vedral}. We exemplify by comparing the distance based measure of entanglement to other well known measures of entanglement for symmetric and non-symmetric Gaussian states produced by sending a two-mode thermal state through lossy optical fibers.

{\textit{Gaussian states.}} Gaussians  continuous variables states 
are standard in quantum mechanics, whose information is stored in two simple quantities: 
the {\it Mean Value Vector} and the {\it Covariance Matrix} (CM) \cite{englert2}. 
Mean values can be displaced by local operations to the null vector,
without affecting entanglement, being usually neglected.  
%
%
%
%
%
For a bipartite system described by bosonic operators, $(a_1,a_2)$, 
the $ 4\times 4$ CM reads, after suitable local operations, as \cite{simon} 
\be                                                                  \label{sf}
{\bf V} = 
\left( \begin{array}{cc}
{\bf V}_1 & \bf C \\
{\bf C}^\dagger & {\bf V}_2
\end{array} \right), 
\,\,{\bf V}_i = n_i \mathbf I , \,\, {\bf C} = 
\left( \begin{array}{cc}
m_s & m_c\\
m_c & m_s
\end{array} \right),
\ee 
$n_i,m_c,m_s \in \mathbb{R}$, %
being hermitian and positive semidefinite, $\textbf V^\dag = \textbf V \ge 0$. 
%
%
Additionally, the non-commutativity of the 
creation/annihilation operators, imposes  a constraint
on $\textbf V$: 
\be
{\bf V} + \frac 12 {\bf E}\ge 0, \label{cond}
\ee
where 
${\bf E}=\text{diag}(\mathbf{Z},\mathbf{Z})$, 
${\bm{Z}}=\text{diag}(1,-1)$. 
Separable Gaussian bipartite states must also obey \cite{simon}%
\be 
{\bf \widetilde V} + \frac 12 {\bf E} \ge 0,  \label{cond2}
\ee
where $ {\bf\widetilde V} = {\bf TVT}$ is achieved by a partial phase 
space mirror reflection,  ${\bf T} = \text{diag}({\bf I},{\bf X})$, 
and ${\bf X} = \text{adiag}(1,1)$. 
It is known that a necessary and sufficient condition for the 
positivity semi-definiteness of a matrix is that its upper left
block be positive definite and the block's Schur complement %
\footnote{The Schur complement of a matrix partitioned as 
           (\ref{sf}) with respect to the upper-left block, 
           say ${\bf V}_1$, is deffined by 
           $\mathcal S(\bf V) := { \bf V}_2  -{\bf C}^\dagger {{\bf V}_1}^{\!\!-1} {\bf C} $, 
           only if ${\bf V}_1$ is not singular.} %
be positive semidefinite. 
Thus the physical positivity criterion (\ref{cond}) applies if and only if \cite{mcol1}
\begin{equation}                                                               \label{condh}
{{\bf V}_1}  + \tfrac 12 {\bf Z} > 0 
\,\,\,\,\,\, \text{and} \,\,\,\,\,\,
\mathcal{ S } ({\bf V} + \tfrac 12 {\bf E})  \ge 0,      
\end{equation}
and the separability condition (\ref{cond2}) holds only if \cite{mcol1}
\begin{equation}                                                               \label{conds}
{{\bf V}_1}  + \tfrac 12 {\bf Z} > 0
\,\,\,\,\,\, \text{and} \,\,\,\,\,\,
\mathcal{ S } ({\widetilde{\bf V} } + \tfrac 12 {\bf E})  \ge 0.                           
\end{equation}
%

\begin{figure}[htbp]
\includegraphics[width=.35\textwidth]{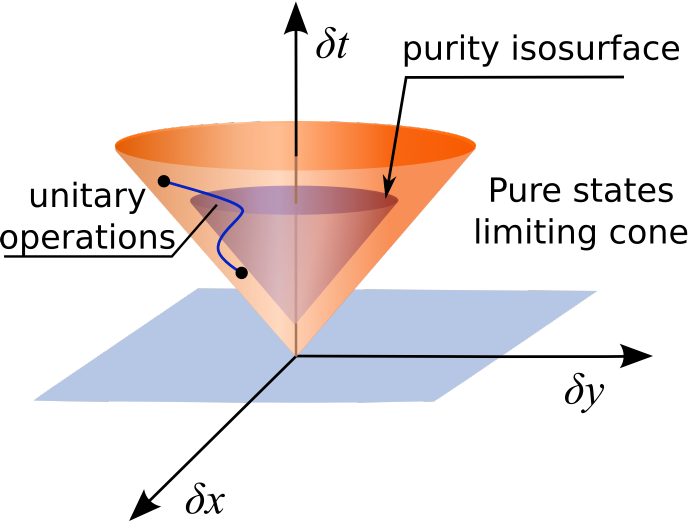}
\caption{(Color Online) Universe cone for any two-mode bipartite GS. 
Unitary operations connect any two-states with same purity lying in a conic 
isosurface of purity.}\label{fig1}
\end{figure}

{\textit{Geometry.}}
In order to explore the geometric features 
of the GS we first write the inequalities in (\ref{condh}) and (\ref{conds}) 
in terms of the matrices entries in (\ref{sf}).
 We verify that the inequalities in (\ref{condh}) reduce to 
the quadratic form 
\be
\delta s^{2}=\delta t^{2} - \delta x^{2} - \delta y^{2} \geq 0,\label{interval}
\ee
having a Minkowski metric, where  
\begin{eqnarray}
\delta {t}^{2} & = & 
                   ( I_1 - \tfrac{1}{4} )^{-1} \left(  I_1 - \tfrac{1}{4}  - 
                       \tfrac{1}{2} I_4 / I_2 \right)^{2} I_2,               \nonumber\\
\delta {x}^{2} & = &       ( I_1 - \tfrac{1}{4} )^{-1} 
                     \left ( \tfrac{1}{4} I_4^2/I_2 - I_1I_3^2 \right) ,     \nonumber \\
\delta {y}^{2} & = &  \tfrac{1}{4}( I_1 - \tfrac{1}{4} )^{-1} 
                   \left( I_1 - \tfrac{1}{4} + I_3  \right)^{2},                    
                                                                             \label{minkphysINV1}
\end{eqnarray}
with the local invariants \cite{simon} being
$ I_1 = \det\mathbf{V} _1$, 
$ I_2 = \det\mathbf{V}_2 $,
$ I_3 = \det\mathbf{C} $, and 
$ I_4 = \text{tr} (\mathbf{V}_1 \mathbf{Z} \mathbf{C} \mathbf{Z} \mathbf{V}_2 
                                \mathbf{Z} \mathbf{C^\dagger} \mathbf{Z} )    $. 
In this $1+2$ dimensional space, a separatrix is defined by $\delta s^2=0$
setting the boundary for discerning physical from nonphysical states. 
States lying at the boundary are pure bipartite GS, corresponding
to equality in (\ref{condh}).
By computing all the terms in (\ref{interval}) we get 
\begin{equation}
 \delta s^{2} =  \det {\bf V} -  
                 \tfrac{1}{4} \sigma_{ { \bf V } } + \tfrac{1}{16},
\end{equation}
where $\det {\bf V} = I_1 I_2 + I_3^2 - I_4 $ and 
$ \sigma_{ {\bf V} } = I_1 + I_2 + 2 I_3$ \cite{serafini}.
An arbitrary pure global state is characterized by 
$\det {\bf V}= 1/16$ and $\sigma_{{\bf V}}=1/2$, 
and so $\delta s^{2} = 0$ and is located at  the external conic boundary, defining an 
isosurface for states with unit purity $\mathcal{P}= \rm{Tr}(\rho^2)=1$ (See Fig. \ref{fig1}). 
Conic isosurfaces  
inside the volume define states with same purity, $\mathcal{P}= 1/(4\sqrt{\det {\bf V}})$. 
Therefore, analogously to intervals in the space-time, defined as 
the distance between two points (events)  in the light-cone, 
an interval here connects a given state with a certain purity $\mathcal{P}<1$ to its closest pure
state situated at the external surface of the  physical cone of existence $\mathcal{P} =1$. 
Since both purity $\mathcal{P}$ and $ \sigma_{ {\bf V} }$ (the seralian) are preserved by unitary operations, 
all states lying in a $\mathcal{P}$ isosurface are connected by unitary operations. 
So the Lorentz invariance of 
$\delta s^{2}$ is associated to invariance of $\mathcal{P}$ under an arbitrary unitary  operation, 
where ${\bf V'}={\bf S}^\dagger {\bf V}{\bf S}$ is the CM under a symplectic transform 
${\bf S}$ over ${\bf V}$, related to the arbitrary unitary operation $U$ by 
$U{\bf v}U^{-1}={\bf S}{\bf v}$: ${\bf v} = (a_1^\dagger, a_1,a_2^\dagger,a_2)^\dagger$.

In relativity the causal structure allows that at any event another light-cone be defined, 
therefore restricting all world lines. For the GS depicted in a hyperbolic space 
(Minkowski picture), 
the $\mathcal{P}=1$ cone defines all states that can be generated from the vacuum 
(as all GS can be generated by convenient Gaussian operations over the vacuum). 
Trace preserving operations may preserve purity (if unitary) or decrease it (if not unitary). 
Being at a certain state of the cone of existence, a new set of Gaussian operations lead to 
any new state inside the cone if {non-unitary}
trace preserving operations are allowed.  
While local unitary operations must connect states in a specific conic isosurface, 
arbitrary (trace preserving) non-unitary operations, can move states from the surface 
to any state inside the cone volume, which in that case preserves (or decreases) 
the amount of entanglement depending on the nature of the operation.  
Here, similarly to the limiting velocity of light in relativity, 
the purity $\mathcal{P}=1$ is the limiting quantity.

{\textit{The squared distance for entanglement.}} 
Global operations can certainly change the amount of entanglement of a given state, transforming from one state to another with a different amount of entanglement. 
However local (non-stochastic) operations cannot change it, while they certainly change the state.
So local operations form a special class of causal operations connecting states 
with the same amount of entanglement. 
Let us discuss this point with an appropriate picture, 
rewriting the inequalities in (\ref{conds}) as
%
%
\be
\delta \tilde{s}^{2} = \delta {t}^{2} -
\delta {x}^{2} - \delta \tilde{y}^{2} \geq 0                                      \label{interval2}
\ee
with
\br                                                                               \label{minksepINV}
\delta\tilde{y}^{2}  & = & \tfrac{1}{4} \left( I_1 - \tfrac{1}{4}\right)^{-1} 
                           \left( I_1 - \tfrac{1}{4} -  I_3  \right)^{2}.
\er
An entangled GS necessarily implies $I_3 < 0$ \cite{simon}. Therefore
Eq. (\ref{interval2}) turns out to be the Simon \cite{simon} separability criteria for GS. 
So the Minkowski structure emerges with a separatrix given by  $\delta \tilde{s}^{2}=0$, 
dividing the space into \textit{separable-like} and \textit{entangled-like} regions. 
$\delta \tilde{s}^{2}\geq 0$ includes all separable states, 
while $\delta \tilde{s}^{2}<0$ corresponds to all entangled GS. 
%

We must understand the meaning of such a relation between both regions, 
and for that we address to Fig. 2. The Minkowski space deals with intervals (between events), 
while the symplectic deals with states. Again we match these two features by identifying 
the meaning of the invariant squared distance interval in (\ref{interval2}). 
\begin{figure}[htbp]
\includegraphics[width=.35\textwidth]{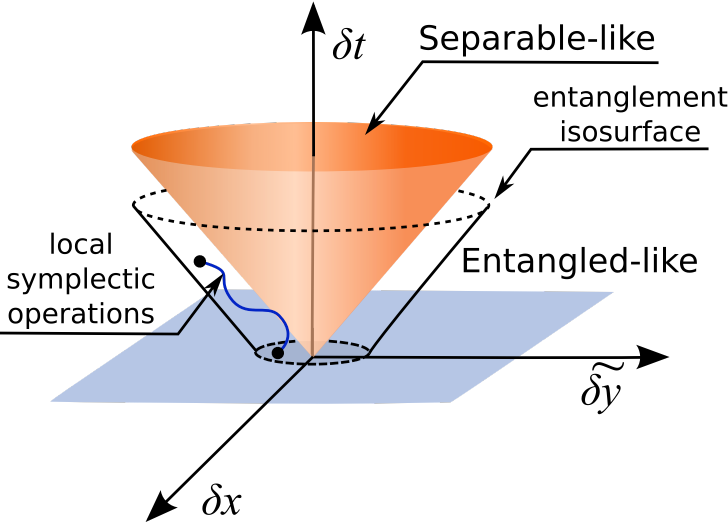}
\caption{(Color Online) Separation between the separable-like and entangled-like regions after partial 
transposition of a two-mode bipartite state. Any two states lying at a conic entanglement isosurface can 
be connected through local ${\rm Sp}( 2 , \mathbb R ) \otimes
{\rm Sp}( 2 , \mathbb R )$ operations, and therefore have the same amount of entanglement.}\label{fig2}
\end{figure} 
The interval defined in the hyperbolic space is actually 
a distance between the given state and the closest separable state lying at the separatrix.
Since entanglement does not change due to local unitary operations, 
the two regions are disconnected by any ${\rm Sp}( 2 , \mathbb R ) \otimes
{\rm Sp}( 2 , \mathbb R )$ unitary operation. In fact only points in the Minkowski 
space which have the same entanglement can be connected by those operations. 
Therefore any two states with the same amount of entanglement belong to the same conic isosurface. 
The Lorentz invariance of $\delta \tilde{s}^{2}$ 
is associated to {the} invariance of entanglement of two-mode bipartite GS under 
arbitrary local symplectic
unitary operations, i.e., for ${\bf V'}={\bf S_L}^\dagger {\bf V}{\bf S_L}$, 
where ${\bf S_L}$ must be  
\be                                                \label{local2}
U_L{\bf v}U^{-1}_L = {\bf S_L} {\bf v},\;\;\; 
{\bf S_L} = \text{diag}({\bf S_1},{\bf S_2}),\ee 
with the condition ${\bf S_L}^{\! \! \!-1}={\bf E S_L}^{ \! \! \dagger}{\bf E}$. 
In a simplified scenario, any state living on the $(y,t)$ 
plane is linked to other states with constant $\delta {t}$ by a rotation in the $(x,y)$ plane. 
At that plane, violating inequality (\ref{interval2})
means that the state lies on a line parallel to the cone's surface: 
$ \delta {\tilde{t}}^2 - \delta { \tilde{y}}^2 = - \delta \tilde{s}^{2}$. 
Since all states with the same $\delta \tilde{s}^{2}$ are equidistant to the separatrix 
they are connected through operations in (\ref{local2}) 
lying in a straight line parallel to the separability 
boundary, $ \delta {\tilde{t}}^2 = \delta { \tilde{y}}^2 $, as in Fig.~\ref{fig2}.

{\textit{Entanglement properties and quantification.}}
Now, we investigate the quality of $|\delta \tilde{s}^{2} |$ 
as a good measure of entanglement, which requires it to satisfy some specific properties \cite{zyczkowski} in the context of GS and 
Gaussian operations \cite{GiedkeCirac,fiurasec}. 
It will be useful for us rewrite eq. (\ref{interval2}) as
\begin{equation}
\delta \tilde{s}^{2} = \det \left({\bf \widetilde V} + \tfrac 12 {\bf E}\right) = 
(\tilde{n}_+^2 - 1/4)( \tilde{n}_-^2 -1/4) ,    \label{sep3}                  
\end{equation}
where $\tilde{n}_\pm$ are the symplectic eigenvalues of $\tilde {\bf V}$, 
explicitly given by \cite{serafini} 
\begin{equation}
\tilde{n}_{\pm}^2 = \tfrac{I_1 + I_2}{2} -  I_3   \pm 
                    \sqrt{ \left( \tfrac{ I_1 - I_2 }{2} \right)^2 -  (I_1 + I_2) I_3 +  I_4 }.   \label{simpeig}                
\end{equation}
Furthermore, $\tilde {\bf V}$ is positive semidefinite 
and $\tilde{n}_+  \ge \tilde{n}_- \ge  1/2$ for a separable state, 
while for an entangled state $ 0 < \tilde{n}_- < 1/2$ fulfilling  
$\delta \tilde{s}^{2} < 0$ (in analogy to the space-like condition in relativity).
Eqs. (\ref{sep3}) and (\ref{simpeig}) link the squared distance 
$\delta \tilde{s}^{2}$ (when $\delta \tilde{s}^{2}\le 0$)  with the Simon separability criteria 
for bipartite GS \cite{simon} expressed as a function of 
the symplectic eigenvalues.
In fact, measuring entanglement by distances in a Hilbert space
(see for instance \cite{paulina3,mcol2} for the Bures metric)
requires a hard minimization procedure over 
a set of separable states. 
Here instead, $\delta \tilde{s}^{2}$ does not require any minimization 
procedure since it is given due to the Minkowski structure as a straight 
line between the two parallel conic surfaces, 
one containing the given state and the second its closest separable state. 
Therefore   $\delta \tilde{s}^{2}$ satisfy the 
{\it computability} requirement.

The {\it discriminance} requirement states that 
$\delta \tilde{s}^{2} = 0$ if and only if $\hat \rho $
is separable, and this is true for all bipartite GS, 
since there is no bipartite GS with bound entanglement \cite{werner}.
Two states living closer inside the existence cone 
have partial transpositions also close to each other 
since by construction 
the difference between the original state and the partially 
transposed is a sign in $I_3$, see eq.(\ref{minksepINV}): 
this defines the {\it asymptotic continuity} for the measure.

Given a convex decomposition of a quantum state, 
the entanglement of this state cannot be less than the convex sum of 
the entanglement of each part of the decomposition. 
Given two arbitrary two-mode bipartite GS, $\hat \rho$ and $\hat \rho'$ with corresponding entanglement $\delta \tilde{s}^{2}$ and $\delta \tilde{s}'^{2}$ then 
\begin{equation}
\hat \rho = \int\! d^2\!\alpha \, d^2\!\beta \,\, P(\alpha,\beta) \,\,
{\hat D}_{\alpha\beta} \, \hat \rho' {\hat D}^\dagger_{\alpha\beta}  
\rightarrow |\delta \tilde{s}^{2}| \le |\delta \tilde{s}'^{2}| ,\label{P}
\end{equation}
where $P$ is a normalized Gaussian probability function with CM $\mathbf P$, and
${\hat D}_{\alpha\beta}$ is the displacement operator \footnote{ 
The implication relation in (\ref{P}) is a consequence of the 
locality of the displacement operator (it does not affect the entanglement) 
and of the unity integration of $P$. }.
To prove the necessary condition of {\it convexity}, given the CMs
of the above relation ${\bf V} =  {\bf P} + {\bf V'}$ we  derive that
\begin{equation}
\left|\det \left({\bf \widetilde V}   + \tfrac 12 {\bf E}\right)\right| \ge 
\left|\det \left({\bf \widetilde V '} + \tfrac 12 {\bf E}\right)\right|.  
\end{equation} 
GS entanglement   cannot be distilled 
by LOCC Gaussian operations\cite{fiurasec,GiedkeCirac}. Therefore any good entanglement measure 
cannot decrease under  
these operations - a property called {\it monotonicity}. 
To prove the monotonicity for $\delta \tilde{s}^{2}$, 
first let us note that all stochastic Gaussian LOCC, represented by 
a $8 \times 8$ CM $\Gamma$ acting on a input GS with CM $\bf V$, 
can be reproduced by means of a deterministic Gaussian LOCC \cite{fiurasec}, 
furthermore $\Gamma$ is separable with respect to the input (with CM $\bf V$) 
and output states (with CM $\bf V'$). 
Under these conditions $\bf V ' \le \bf V$ implying necessarily  \footnote{Remark on Eq. (17) in \cite{GiedkeCirac} 
and set the extremal case $p = 1$. Note also that the symbol 
$V$ in this reference corresponds to the entanglement 
measure not to the covariance matrix.}
that 
$ {\bf \widetilde V '} \le {\bf \widetilde V }$ \cite{GiedkeCirac}.      
Is is direct to see that 
$\left|\det \left({\bf \widetilde V ' }  + \tfrac 12 {\bf E}\right)\right| \le
\left|\det \left({\bf \widetilde V   }  + \tfrac 12 {\bf E}\right)\right|$.  
All those properties guarantee that $ | \delta \tilde{s}^{2}| $ (when $\delta \tilde{s}^{2}\le0$) is an entanglement monotone \cite{zyczkowski}\footnote{Remark that it is meaningless 
to use $|\delta \tilde{s}^{2}|$ as a measure when 
$\delta \tilde{s}^{2} > 0$, since in this circumstance the state 
is separable.}.

It is interesting to compare the Minkowski interval $\delta \tilde{s}^{2}$ 
with other available measures of entanglement. For that we define 
%
%
\begin{equation}                                                                         \label{efg}
{\rm E}(\rho_{12}) = f \left( 2 \sqrt{ \frac{\delta \tilde{s}^{2} }
                                 { (\tilde{n}_+^2 - 1/4) } 
                                 + \frac{1}{4}  }  \right),  
\end{equation}
being $f(x)$  a monotonically decreasing function 
over the interval $x \in (0,1]$ \footnote{
Note that once one determines $\tilde{n}_-$, 
$\tilde{n}_+$ will automatically be defined by (\ref{simpeig}).}. 
In that form Eq. (\ref{efg}) can be connected to two 
distinct entanglement measures -
the Logarithmic Negativity (LN) \cite{serafini} and the Entanglement of Formation EoF for symmetric GS \cite{Gie03}. The LN measure is given by taking $f(x) = - {\rm ln }(x) $ in Eq. (\ref{efg}) 
For symmetric GS ($I_1 = I_2$),  
the EoF can be computed analytically \cite{Gie03}, and is given by taking $f(x) = c_+(x) \log_2(c_+(x)) - c_-(x)\log_2(c_-(x))$ with
$c_{\pm}(x)=(x^{-1/2}\pm x^{1/2})^2/4$ in Eq. (\ref{efg}).
Both the LN and the EoF are monotonically decreasing function of $\tilde{n}_-$:  
the closer $\tilde{n}_-$ is to zero, the state is more entangled.      There is no closed analytical 
expression for the EoF for non-symmetrical
GS \cite{paulina2}, whose computation relies on a minimization procedure 
 \cite{wolf}.  
%
%
%
We employ this same formula to calculate a lower bound for the EoF for non-symmetric GS \cite{rigolin}.
%

\begin{figure}[!ht]
\includegraphics[width=.4\textwidth]{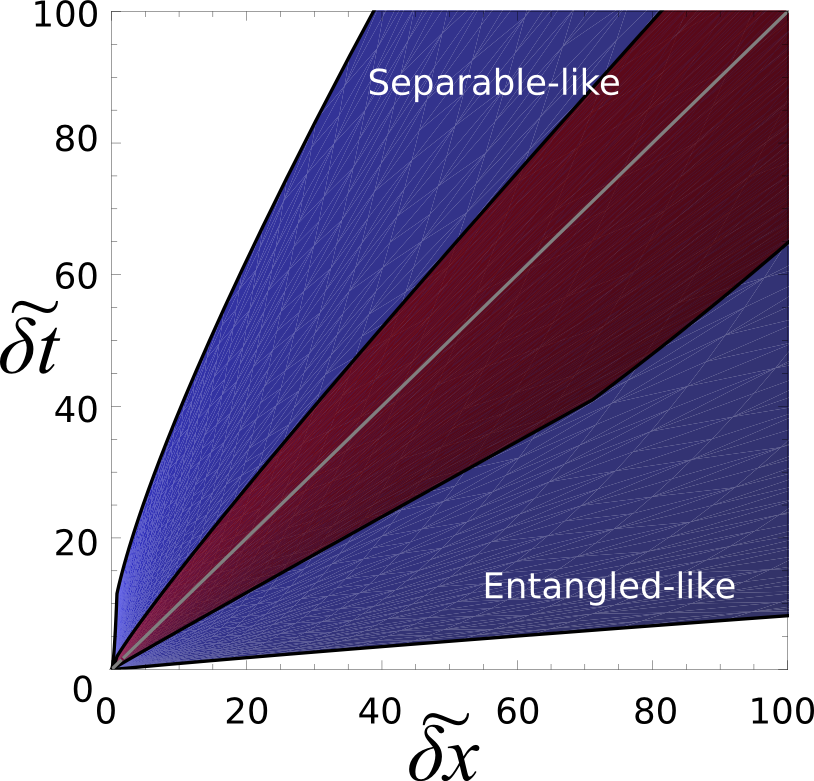}
\caption{ (Color Online) 
Blue Region: 
Set of all TMTSS with a fixed 
dissipative parameter of $d=2.5$ and varying thermal 
photon number $\bar n$ from $0$ to $1.5$, 
and squeezing rate $r$ from $0$ to $3.0$. 
The color map in the (blue) limited region indicates 
from light to darker the increasing feature of $r$. 
Red Region: 
Set of all TMTSS above after asymmetric action of a lossy 
fiber with $\ell = 0.5$. The effect of the asymmetry is to constrain the available states to a  smaller area around the separatrix.
%
}            \label{fig3}
\end{figure}

We now concentrate on the 
kind of GS actually generated experimentally ---
the two-mode thermal squeezed state (TMTSS) \cite{daffer} ---  
produced in a nonlinear crystal with internal noise. 
These states are characterized by the following 
values for the parameters: 
$ n \equiv n_1 = n_2 =  (h_1 + h_2)/4$,
$m_s = 0 $, and $ m_c = (h_1 - h_2)/4 $, 
with $h_i  =  \left\{ e^{-p_i} + d \, (2\bar n + 1)
          \left[ { (1-e^{-p_i}) }/{p_i} \right] \right\}$ and
$ p_1 = d + 2r $ and $ p_2 = d - 2r $. 
$ d$
is a dissipative parameter,
and $r$
 is the squeezing
parameter. $\bar n$ is the mean number of thermal photons
introduced by the quantum noise.
Therefore 
$\delta{t}^{2}= n^2 \left( n^2-\tfrac{1}{4}-{m_c^2} \right)^{2}/(n^2-\tfrac{1}{4})$, 
$\delta\tilde{y}^{2}=\tfrac{1}{4}\left(n^2-\tfrac{1}{4}+m_c^2\right)^{2}/(n^2-\tfrac{1}{4})$
and $\delta{x}^{2}=0$.
The 
measure (\ref{efg}) turns out to be simply
${\rm E} ( \rho_{12} ) = f \left[ 2 (n - |m_c| ) \right]$.
It vanishes at the separability boundary 
$ m_c = \pm ( n - 1/2 ) $. 
%
Since any ${\rm Sp}( 2 , \mathbb R ) \otimes {\rm Sp}( 2 , \mathbb R )$ 
unitary operation does not change the amount of entanglement, 
necessarily all states connected through 
it are located on lines parallel to the separatrix (see Fig. 3). 
For a fixed $ d = 2.5 $
 as $r$ is  increased
the state gets more entangled, while by increasing $\bar n$ 
it tends to lie on the separable-like region.
Asymmetry effects can be introduced by
assuming that the TMTSS is distributed by lossy optical fibers \cite{wolf}. 
The fibers output field state will have a CM of the form (\ref{sf}) 
with
$n'_i \equiv (n_i - 1/2) T_i^2 + 1/2$, for $i = 1,2$  and 
$m'_c \equiv m_c T_1 T_2$. 
The transmission coefficients in the 
asymmetric configuration are $T_1 = 1 $, $T_2 = \exp(-\ell)$ \footnote{Note that the symmetric configuration
corresponds to $ \ell = 0$.}, 
where $\ell$ is a dimensionless length related 
to the fiber's absorption. 
Now  
$\delta {t}^{2}={n'_2}^2\left({n'_1}^2-\tfrac{1}{4}-\tfrac{n'_1}{n'_2}{m'_c}^2 \right)^{2}/(n_1' - \tfrac{1}{4})$,
$\delta \tilde{y}^{2} =  \tfrac{1}{4}\left({n'_1}^2-1/4+{m'_c}^2\right)^{2}/(n_1' - \tfrac{1}{4})$, 
$\delta{x}^{2}=0$
and the separatrix will be at $(n'_1 \pm {1}/{2})(n'_2 \pm {1}/{2}) = {m'_c}^2$. 
In Fig. \ref{fig3}, 
we see that due to the additional noise introduced by the fiber, 
the states are confined to a region around the separatrix.

\begin{figure}[!ht]
\includegraphics[width=.45\textwidth]{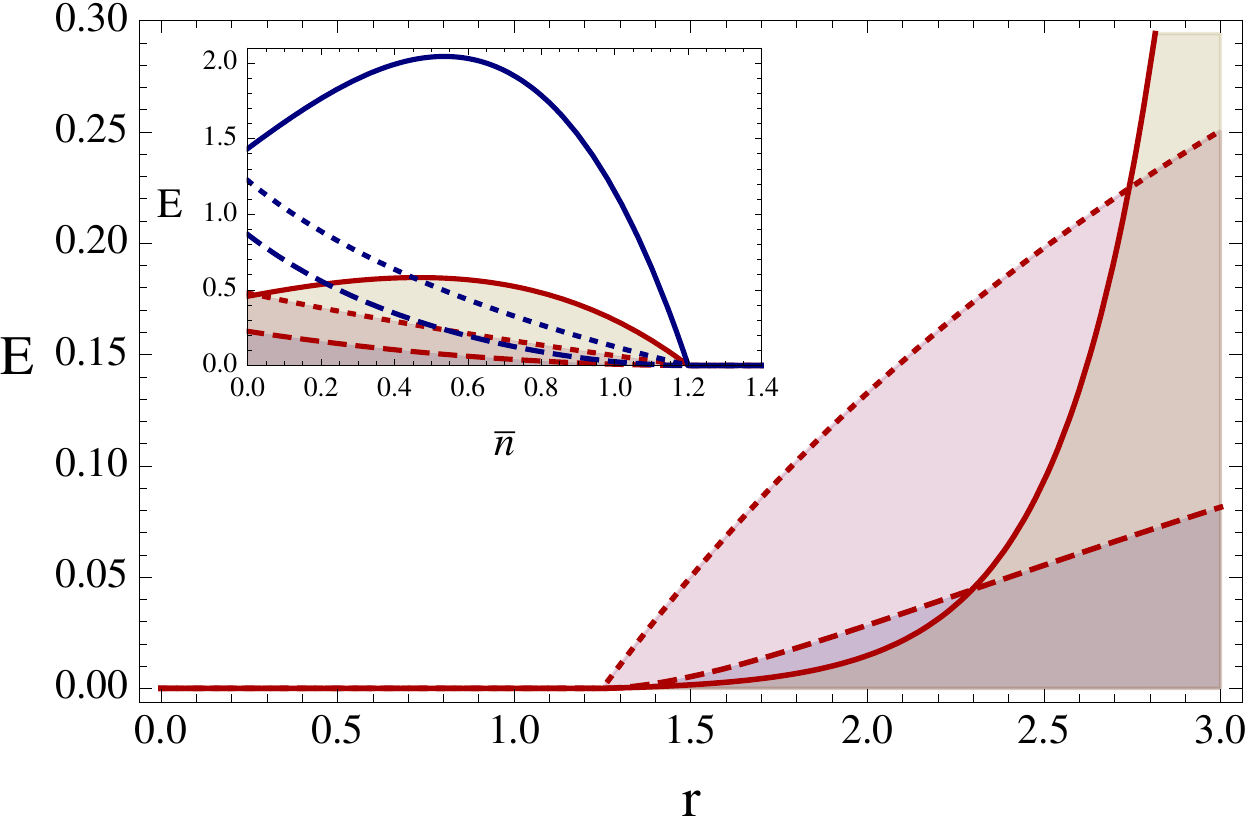}
\caption{ 
(Color Online) Measuring entanglement of (asymmetric) GS 
living in the red region of Fig.\ref{fig3} using the 
Minkowski Distance $|\delta s^2|/2000$ (continuous lines), 
the EoF bound (dashed lines) 
and the LN (dotted line) as a function of $r$ with fixed $d=2.5$, $\bar{n} = 0.5$ and $\ell = 0.5$. 
In the inset we show the same quantities for a symmetric 
GS ($\ell = 0$) living in the blue region of fig.\ref{fig3} 
and for the asymmetric GS (shadowed curves) with $\ell = 0.5$ 
living in the red region of fig.\ref{fig3} as a function of the 
mean thermal number $\bar{n}$ with $r = 3$ and $d = 2.5$.
}
\label{fig4}
\end{figure}

%
%

%
To  compare the different measures, 
we plot $|\delta \tilde{s}^{2}|$ in Fig. \ref{fig4} 
and the two cases for (\ref{efg}): 
the EoF bound 
and the LN when $r$ and $\bar n$ increase.  
The measures given by (\ref{efg}) have qualitatively 
the same behavior (with the LN being always greater than the  EoF bound) 
for symmetric and asymmetric states. On the other side, 
$|\delta \tilde{s}^{2}|$ is always greater then both 
(note that this function is rescaled in Fig.\ref{fig4}).
As $r$ increases from zero to $r_0 \approx 1.25$, the noise and dissipation of the crystal  
are responsible for the separability of the TMTSS. After this threshold the state becomes entangled as can be seen for the three plotted functions.
The behavior with $\bar n$ variation is shown in the inset and now the measures differ qualitatively: 
as the functions (\ref{efg}) always decrease, 
the distance reaches a maximum value and then decreases to zero.

{\textit{Discussion.}} We have 
explored the symplectic and Lorentz groups relation to 
investigate {some formal analogies} with special relativity,
{related to quantum mechanical features of GS as}
purity and entanglement. 
{Particularly, we have observed that a monotone 
distance based entanglement measure can be analytically given, being the 
optimization, usually required for this kind of measure, directly given 
by the Minkowski structure.}
We remark that the present description can be generalized to include
non-Gaussian states as well. In that situation there are states, 
which are entangled although satisfying $\delta \tilde{s}^{2}\ge 0$, 
thus lying within the cone. Those states are not detected by the 
PPT criterion, and are known as bound entangled states. So, what is mostly interesting 
in the Minkowski diagram in Fig. 2 is that it then splits the space into a region 
containing only entanglement that can be distilled (by non-Gaussian operations), 
and a region containing separable states and entangled states that cannot be 
distilled by any kind of local operations.
Finally we suggest that beyond the clear importance of this picture for entanglement quantification, given high degree of  
control in the experimental generation Gaussian quantum light fields, 
one could think of this system as a general analog simulator for relativistic phenomena.

%

\begin{acknowledgements}
{This work is supported by the Brazilian funding agencies CNPq 
and FAPESP through the Instituto Nacional de Ci{\^e}ncia e 
Tecnologia - Informa\c{c}{\~a}o Qu{\^a}ntica (INCT-IQ).
F.N. wishes to thank financial support from FAPESP (Proc.2009/16369-8).}
\end{acknowledgements}

\end{document}